\documentclass[12pt]{article}
\usepackage{latexsym} \usepackage{epsfig}
\usepackage{axodraw}
\begin{document}

  \title{\bf Conformal Invariance = Finiteness\\[0.3cm]  and Beta Deformed N=4 SYM Theory}
\author{D. I. Kazakov$^{1,2}$ and  L. V. Bork$^{2,3}$}
\date{}
\maketitle \vspace{-0.8cm}
\begin{center}
$^1$ Bogoliubov Laboratory of Theoretical Physics, Joint
Institute for Nuclear Research, Dubna, Russia, \\
$^2$Institute for Theoretical and Experimental Physics, Moscow, Russia, \\
$^3$ Moscow Engineering Physics Institute, Moscow, Russia.
\end{center}

\begin{abstract}
We claim that if by a choice of the couplings the theory can be made conformally
invariant (vanishing of the beta functions) it is automatically finite and vice versa.
This is demonstrated by explicit example in supersymmetric gauge theory. The formalism is
then applied to the beta deformed ${\cal N}=4$ SYM theory and it is shown that the
requirement of conformal invariance = finiteness can be achieved for any complex
parameter of deformations.
\end{abstract}

\section{Introduction}

The ${\cal N}=4$ supersymmetric Yang-Mills theory (SYM) attracts much attention these
days providing the playground to test nonperturbative features of quantum field theory.
This is related to the property of conformal invariance which is unique for four
dimensional field theories~\cite{fin}. Another remarkable feature of the ${\cal N}=4$ SYM
theory is that via the AdS/CFT correspondence~\cite{ads} it is related to a supergravity
theory and one can get deeper understanding of duality between these two theories.
Combined information may lead to new insight in gauge theories beyond the usual PT.

Note that the above mentioned AdS/CFT correspondence  requires from the field theory to
be conformally invariant and not necessarily obtaining the full ${\cal N}=4$
supersymmetry. From this point of view it would be interesting to consider the other
conformally invariant theories and to find the corresponding supergravity backgrounds. Of
special interest is a marginally deformed ${\cal N}=4$ theory analyzed in~\cite{deform}
for which the supergravity dual description has been found in~\cite{LM}. This the
so-called $\beta$ deformation of the original ${\cal N}=4$ SYM theory has been studied
in~\cite{Z,Z1} with the aim to get the conditions for its finiteness and conformal
invariance. The authors performed a thorough analysis of the UV divergences in the
framework of dimensional regularization (reduction) and found out that one can reach the
desired goal if the deformation parameter $\beta$ is real. They also claim that the
requirements of finiteness and conformal invariance are not simultaneously satisfied and
if one requires only conformal invariance to be valid then the complex values of $\beta$
are also allowed~\cite{Z1}. This problem has been also considered in~\cite{RSS} where it
was shown that conformal invariance understood as vanishing of the beta function holds in
all orders of PT for any complex value of the deformation parameter provided one properly
adjusts the couplings.

The aim of this paper is to show that the above mentioned mismatch between conformal
invariance and  finiteness is a result of mistreatment of dimensional regularization
(reduction). If applied properly, one can reach both  conformal invariance and finiteness
simultaneously, thus allowing for the complex $\beta$ deformations. Moreover, one can
construct the whole family of conformally invariant and finite ${\cal N}=1$ SYM theories,
however, their dual description is not known so far.

\section{The General Formalism}

The problem of finiteness in SYM theories has been studied long time ago~\cite{finite}
and the formalism has been developed~\cite{K1} that allows one to treat the theory within
the dimensional regularization (reduction). For completeness we briefly summarize  it
below.

Let us consider an ${\cal N}=1$ SYM theory formulated in terms of ${\cal N}=1$
superfields with an arbitrary cubic superpotential containing some set of Yukawa
couplings $\{y\}$. We assume that a theory is gauge invariant and for simplicity consider
the background gauge. Then from the non-renormalization theorems~\cite{non-ren} one gets
that in the chiral sector only the propagators are divergent and have to be renormalized
while the vertices are finite. As for the gauge sector, in background gauge the
renormalization of the vertex coincides with that of the gauge propagator, so one can
also consider the gauge propagator only~\cite{back}.

At the one loop order to get the gauge propagator finite one has to make the proper
choice of the matter superfields. The following requirement is to be
satisfied~\cite{1loop}:
\begin{equation}\label{trace}
\sum_R T(R)=3 C_2(G),
\end{equation}
where $T(R)$ is the Dynkin index of a given representation $R$ and $C_2(G)$ is the
quadratic Casimir operator of the group.

Provided the requirement (\ref{trace}) is satisfied the only divergence one should take
care of is the one of the chiral field propagator. This is a consequence of the following
theorem~\cite{Theorem}:

{\it Theorem}: If ${\cal N}=1$ supersymmetric gauge theory is finite in $L$ loops, the
gauge propagator is finite $L+1$ loops.

The same statement follows also from the explicit expression for the gauge beta function
written in terms of the anomalous dimensions of the chiral fields in some particular
scheme~\cite{NSVZ}
\begin{equation}\label{beta}
  \beta_g=g^2\frac{\sum T(R)-3C(G)-\sum T(R)\gamma(R)}{1-2gC(G)}, \ \ \ \ g\equiv
  g^2/16\pi^2.
\end{equation}

Thus, if the anomalous dimensions of the chiral fields vanish, so does the gauge and
Yukawa beta functions and the theory is conformally invariant. In some other gauges (for
instance in components) one can have non-vanishing anomalous dimensions of some fields or
vertices, but the beta functions still vanish. This situation is also attributed to
conformal invariance since only the gauge invariant quantities make sense. In what
follows we will assume the simplest possibility when all anomalous dimensions vanish and
will call this situation  conformal invariance.

Now the question is: how to reach this goal, i.e. how to get conformal invariance? And
the related one: is the theory finite (that is all divergences cancel) in this case? We
show below how it may be done in the framework of dimensional regularization (reduction)
and give a positive answer to the second question. First, we analyse both problems
(conformal invariance and finiteness) separately and then show that this is the same.

To study conformal invariance or vanishing of the anomalous dimensions one has first to
apply some regularization and some renormalization scheme. In general the anomalous
dimensions are scheme dependent but if they vanish, they vanish in any scheme. We adopt
dimensional regularization or more precisely dimensional reduction~\cite{DR} since
dimensional regularization does not support supersymmetry. We ignore the problem of
inconsistency of dimensional reduction in higher orders~\cite{scope} assuming that it is
adjusted by finite corrections. We adopt also the $\overline{MS}$ renormalization scheme.
Then the chiral field renormalization constant has the form
\begin{equation}\label{zet}
  Z_{2i}^{-1}=1+\sum_{n=1}^{\infty}\frac{C_n^{(i)}(\{y\},g)}{\varepsilon^n}, \ \ \
  C_n^{(i)}(\{y\},g)=
  \sum_{k=n}^\infty C^{i}_{kn}(\{y\},g),
\end{equation}
where the coefficient functions $C^{i}_{kn}(\{y\},g)$ are the homogeneous polynomials in
$y_i$ and $g$ of the order of $k$.\footnote{Hereafter for the shorthand notation we use
$g=g^2/16\pi^2,y_i=y_i^2/16\pi^2$.} The anomalous dimensions $\gamma_i$ depend on
renormalized couplings $\{y\}$ and $g$ and are given by the single pole terms
\begin{equation}\label{anom}
  \gamma_i(\{y\},g)=\sum_k k C^{i}_{k1}(\{y\},g).
\end{equation}
In the lowest orders one has
\begin{equation}\label{anom2}
\gamma_i(\{y\},g)=  B^i_{1j}y_j+B^i_{10}g+\sum_{j,k} B^i_{2jk}y_jy_k+
  \sum_j B^i_{2j}y_jg+B^i_{20}g^2+....,
\end{equation}
where $B^i_{j..}$ are some numbers.

The vanishing of anomalous dimensions can be achieved by choosing the Yukawa couplings in
the form of perturbation series over $g$~\cite{finite}
\begin{equation}\label{yuk}
  y_i=\alpha_{0i}^{(0)} g+ \alpha_{1i}^{(0)} g^2+ \alpha_{2i}^{(0)} g^3+...,
\end{equation}
where the coefficients $\alpha_{ni}^{(0)}$ are calculated order by order in PT solving
the system of linear algebraic equations. To guarantee the existence of solution the
one-loop matrix $B^i_{1j}$ has to be non-degenerate. This has to be explicitly checked in
one loop. Then the procedure works in all orders.

This is not enough, however, to cancel all the pole terms in $Z$ factors (\ref{zet}). At
the same time  finiteness of $Z$ would mean the finiteness of a theory. Indeed, to cancel
the pole terms one has to write down eq.(\ref{yuk}) for $\varepsilon\neq 0$ which means
that one has a double series~\cite{K1}
\begin{eqnarray}
  y_i&=&g\left(\alpha_{0i}^{(0)}+\alpha_{0i}^{(1)}\varepsilon+ \alpha_{0i}^{(2)}\varepsilon^2+...+
   \alpha_{0i}^{(n-2)}\varepsilon^{n-2}+\alpha_{0i}^{(n-1)}\varepsilon^{n-1}+ \alpha_{0i}^{(n)}\varepsilon^n+...\right) \nonumber\\
  &+&g^2\left(\alpha_{1i}^{(0)}+\alpha_{1i}^{(1)}\varepsilon+ \alpha_{1i}^{(2)}\varepsilon^2+...+
  \alpha_{1i}^{(n-2)}\varepsilon^{n-2}+\alpha_{1i}^{(n-1)}\varepsilon^{n-1}+...\right) \nonumber\\
  &+&g^3\left(\alpha_{2i}^{(0)}+\alpha_{2i}^{(1)}\varepsilon+ \alpha_{2i}^{(2)}\varepsilon^2+...+
  \alpha_{2i}^{(n-2)}\varepsilon^{n-2}+...\right) \nonumber\\
&+& ................  \nonumber \\
&+&g^{n-1}\left(\alpha_{n-2i}^{(0)}+\alpha_{n-2i}^{(1)}\varepsilon+
...\right) \nonumber\\
&+&g^{n}\left(\alpha_{n-1i}^{(0)}+...\right) .\label{yuk2}
\end{eqnarray}
In a given order of PT equal $n$ one needs all terms of the double expansion with a total
power of $g\cdot \varepsilon$ equal $n$. The existing freedom of choice of the
coefficients $\alpha_{ki}^{(m)}$ is enough to get {\it simultaneously} the vanishing of
the anomalous dimensions (read {\it conformal invariance}) and the pole terms in $Z$
factors (read {\it finiteness}). The coefficients from $\alpha_{ni}^{(0)}$ to
$\alpha_{0i}^{(n)}$ calculated in n-th order of PT are related. One can not put either of
them to zero in an arbitrary way.

Notice, however, that if the renormalization constants $Z_i$ are finite, there is no need
to any renormalization at all. One can proceed with the unrenormalized expressions. To
show this we again consider the chiral propagators. Consider the bare chiral propagator
prior to any renormalization given by perturbative expansion (D-algebra had already been
performed)
\begin{eqnarray}
  D_{iB}(\{y_B\},g_B,p^2,\varepsilon)=\label{prop}&&\\&&\hspace*{-5cm}=
  1+\sum_{n=1}^{\infty}\frac{1}{(p^2)^{n\varepsilon}}\left(
  \frac{d_n^i(y_B,g_B)}{\varepsilon^n}+\frac{d_{n-1}^i(y_B,g_B)}{\varepsilon^{n-1}}+...+
  \frac{d_1^i(y_B,g_B)}{\varepsilon}+d_0^i(y_B,g_B)+...\right).\nonumber
\end{eqnarray}

The finiteness now means that all $d_n^i(y_B,g_B), \ n>0$ vanish. It is possible to
achieve this goal without any preliminary renormalization in terms of the bare couplings.
The bare couplings, contrary to the renormalized ones, do not depend on the
renormalization scheme but on regularization. In case of a finite theory they are finite
and related to the renormalized ones by finite renormalization which is scheme dependent.

The coefficient functions $d_n^i(y_B,g_B)$ are also the homogeneous polynomials over
$y_B$ and $g_B$ and to achieve the vanishing of them one can choose the bare Yukawa
couplings in the form of one fold $\varepsilon$ expansion with positive powers of
$\varepsilon$~\cite{K1}
\begin{equation}\label{eps}
  y_{iB}=g_B(\alpha_{0i}^{(0)}+\alpha_{0i}^{(1)}\varepsilon+\alpha_{0i}^{(2)}\varepsilon^2+...).
\end{equation}
The coefficients $\alpha_{0i}^{(n)}$ like $\alpha_{ni}^{(0)}$ above are calculated order
by order of PT again solving the system of linear algebraic equations. In one loop this
system of equations coincides with the one for determining the coefficients
$\alpha_{ni}^{(0)}$ with modified r.h.s. and is solvable if the one loop matrix
$B^i_{1j}$ is not-degenerate. This requirement again guarantees the solution in all
orders. Notice that the vanishing of the simple pole automatically leads to the vanishing
of the higher order poles. This is the consequence of local renormalizability of quantum
field theory.

One can see that the problem of finiteness is easier to address in terms of the bare
quantities. Eq.(\ref{eps}) contrary to (\ref{yuk2}) is linear with respect to $g_B$ and
is easier to implement. But both the ways lead to the same statement: if the theory is
finite it is conformally invariant and vice versa.

\section{Example}

To demonstrate how the above mentioned statements are explicitly realized in the
framework of dimensional regularization (reduction) we consider a toy example which
imitates the situation in beta deformed ${\cal N}=4$ SYM theory.

Let us assume that we have a supersymmetric gauge theory with only one Yukawa coupling
$y$ corresponding to a triple interaction. Consider the propagator of a chiral superfield
calculated up to three loops ($D$ algebra had already been performed)
\begin{eqnarray}
  D_{B}(p^2,g_{B},h_{B})&=&1+(\frac{d_{11}}{\varepsilon}+d_{10}+d_{1-1}\varepsilon)
  \frac{1}{(p^2)^\varepsilon}+
  (\frac{d_{22}}{\varepsilon^2}+\frac{d_{21}}{\varepsilon}+d_{20})
  \frac{1}{(p^2)^{2\varepsilon}}\nonumber \\ & +
&  (\frac{d_{33}}{\varepsilon^3}+\frac{d_{32}}{\varepsilon^2}
  +\frac{d_{31}}{\varepsilon})\frac{1}{(p^2)^{3\varepsilon}}+..., \label{1}
\end{eqnarray}
where the coefficient functions $d_{ij}=d_{ij}(g_B,y_B)$ depend on the bare couplings and
are the homogeneous polynomials of the order $i$.

The renormalization constant which makes the propagator finite in the $\overline{MS}$
scheme is
\begin{equation}\label{2}
  Z_2^{-1}=1+\frac{c_{11}}{\varepsilon}+
  (\frac{c_{22}}{\varepsilon^2}+\frac{c_{21}}{\varepsilon})+
  (\frac{c_{33}}{\varepsilon^3}+\frac{c_{32}}{\varepsilon^2}
  +\frac{c_{31}}{\varepsilon})+...,
\end{equation}
where the coefficients $c_{ij}=c_{ij}(g,y)$ depend on the renormalized  couplings and are
also the homogeneous polynomials of the order $i$.  This expression allows one to define
the anomalous dimension $\gamma$
\begin{equation}\label{3}
  \gamma(g,y)=c_{11}+2c_{21}+3c_{31}+...
\end{equation}
and the Yukawa beta function
\begin{equation}\label{4}
  \beta_y(g,y)=3y\gamma(g,y).
\end{equation}
The bare coupling $y_B$ and the renormalized one are related by
\begin{equation}\label{5}
  y_B=yZ_2^{-3},
\end{equation}
where $Z_2^{-1}$ is given by (\ref{2}). Similarly for the gauge coupling one has
\begin{equation}\label{6}
  g_B=gZ_g,
\end{equation}
where we define
\begin{equation}\label{7}
  Z_g=1+\frac{a_{11}}{\varepsilon}+
  (\frac{a_{22}}{\varepsilon^2}+\frac{a_{21}}{\varepsilon})+ ...,
\end{equation}
and the gauge beta function is
\begin{equation}\label{8}
  \beta_g(g,y)=a_{11}+2a_{21}+...
\end{equation}
For our purposes we will need it up to two loops.

Not all of these coefficients are independent. By pole equations~\cite{pole} the
coefficients of the higher order poles in $Z$ factors can be expressed in terms of the
single pole ones as
  \begin{eqnarray}
    a_{22} & =& \frac 12 \left[a_{11}a_{11}+a_{11}g\frac{da_{11}}{dg}+
    3c_{11}y\frac{da_{11}}{dy}\right],\label{pole}\\
    c_{22} & =&\frac 12 \left[c_{11}c_{11}+a_{11}g\frac{dc_{11}}{dg}+
    3c_{11}y\frac{dc_{11}}{dy}\right] , \nonumber \\
    c_{33} &=& \frac 13 \left[c_{11}c_{22}+a_{11}g\frac{dc_{22}}{dg}+
    3c_{11}y\frac{dc_{22}}{dy}\right] ,\nonumber \\
    c_{32} &=& \frac 13 \left[c_{11}c_{21}+2c_{21}c_{11}+a_{11}g\frac{dc_{21}}{dg}+
    2a_{21}g\frac{dc_{11}}{dg}+3c_{11}y\frac{dc_{21}}{dy}+6c_{21}y\frac{dc_{11}}{dy}\right].\nonumber
  \end{eqnarray}
Moreover, from the requirement that
\begin{equation}\label{9}
  Z_2^{-1}D_B(p^2,g_B,y_B)= \mbox{finite when} \ \varepsilon\to 0,
\end{equation}
where for $g_B$ and $y_B$ one has to substitute expansions (\ref{5},\ref{6}), one finds
the relations between the coefficients $d_{ij}$ and $c_{ij}$. They are
 \begin{eqnarray}
    d_{11} & =& -c_{11} ,\\
    d_{22} & =& c_{22} , \nonumber \\
    d_{21} &=& -c_{21}-c_{11}d_{10}-a_{11}g\frac{dd_{10}}{dg}-3c_{11}y\frac{dd_{10}}{dy},\nonumber \\
    d_{33} &=& -c_{33}, \nonumber \\
    d_{32} &=& -c_{32}-c_{11}d_{21}-d_{11}c_{21}-d_{10}c_{22}-a_{11}g\frac{dd_{21}}{dg}-
   3c_{11}y\frac{dd_{21}}{dy}-c_{11}a_{11}g\frac{dd_{10}}{dg}, \nonumber \\ &&- 6c_{11}c_{11}y\frac{dd_{10}}{dy}
 - a_{22}g\frac{dd_{10}}{dg}-3c_{22}y\frac{dd_{10}}{dy}-
 a_{21}g\frac{dd_{11}}{dg}-3c_{21}y\frac{dd_{11}}{dy}, \nonumber \\
     d_{31}&=&-c_{31}-c_{11}d_{20}-d_{10}c_{21}-d_{1-1}c_{22}-a_{11}g\frac{dd_{20}}{dg}-
   3c_{11}y\frac{dd_{20}}{dy}-c_{11}a_{11}g\frac{dd_{1-1}}{dg}, \nonumber \\
   &&- 6c_{11}c_{11}y\frac{dd_{1-1}}{dy}-a_{21}g\frac{dd_{10}}{dg}-3c_{21}y\frac{dd_{10}}{dy}
 - a_{22}g\frac{dd_{1-1}}{dg}-3c_{22}y\frac{dd_{1-1}}{dy}, \nonumber
  \end{eqnarray}

Having all these expressions one can demonstrate how the cancellation of divergences and
nullification of the beta function work. To imitate the situation in beta deformed ${\cal
N}=4$ SYM theory we take the following expressions for the independent coefficient
functions $d_{ij}$ and $a_{ij}$
\begin{eqnarray}
    d_{11} & =& d_1(g-y) , \label{10}\\
    d_{10} & = & d_0(g+y), \nonumber \\
    d_{1-1} &= & d_{-1}(g+y), \nonumber \\
    d_{21} & =& d_2(g^2+gy+y^2) , \nonumber \\
    d_{20} &= & d_{-2}(g^2+gy+y^2), \nonumber \\
    d_{31} &=& d_3g^3\ \ \mbox{for}\ y=g ,\nonumber \\
    a_{11} &=& 0, \nonumber \\
    a_{21} &=& a_2g(g-y). \nonumber
  \end{eqnarray}
The explicit form of $d_{10},d_{1-1},d_{21}$ and $d_{20}$ is not essential. What is
important they do not vanish at $y=g$. For $d_{31}$ we only need its value at $y=g$.
Eq.(\ref{10})
 means that for the chiral propagator the UV divergence disappears for $y=g$ in the one
loop order, but it does not disappear in two and three loops  (in the real beta deformed
${\cal N}=4$ SYM theory in the planar limit it disappears in 1, 2 and 3 loops~\cite{RSS}
but does not disappear in 4 and 5 loops~\cite{Z})). At the same time the gauge beta
function identically vanishes in one loop and vanishes in two loops for $y=g$ (in the
real beta deformed ${\cal N}=4$ SYM theory it vanishes up to 4 loops for $y=g$).

Given eq.(\ref{10}) one can find the remained coefficient functions. They are
\begin{eqnarray}
    a_{22} & =& 0 ,\\
    c_{11} & =& d_1(y-g) , \nonumber \\
    c_{21} &=& -d_2(g^2+gy+y^2)-d_0d_1(y+g)(y-g)-3d_0d_1y(y-g) ,\nonumber \\
    c_{22} &=&d_{22}=\frac 12 d_1^2(y-g)(4y-g), \nonumber \\
    c_{31} &=& -d_3g^3+15d_0d_2g^3 \ \ \mbox{for}\ y=g , \nonumber \\
    c_{32}&=& -2d_2d_1y(y^2+yg+g^2)-d_2d_1(y-g)(3y^2+2yg+g^2)+2/3a_2d_1g^2(y-g)\nonumber \\
    &-&d_1^2d_0(y-g)(20y^2-4yg-g^2), \nonumber \\
    d_{32}&=& -d_2d_1y(g^2+gy+y^2)-d_2d_1(y-g)(5y^2+3gy+g^2)+1/3a_2d_1g^2(y-g)\nonumber \\
    &&-d_1^2d_0(y-g)(10y^2-2gy-1/2g^2),\nonumber \\
    c_{33}&=&-d_{33}=\frac 16 d_1^3(y-g)(28y^2-20yg+g^2). \nonumber
  \end{eqnarray}

Now one can calculate the anomalous dimension $\gamma$ according to eq.(\ref{3})
\begin{equation}\label{12}
  \gamma=d_1(y-g)-2d_2(g^2+yg+y^2)-2d_0d_1(y-g)(4y+g)-3d_3g^3+45d_0d_2g^3+...
\end{equation}
Vanishing of $\gamma$ can be achieved if one chooses the renormalized Yukawa coupling $y$
in the form of perturbative expansion over $g$ (see eq.{\ref{yuk}))
\begin{equation}\label{13}
  y|_{\varepsilon=0}=g+\alpha_1^{(0)}g^2+\alpha_2^{(0)}g^3+...
\end{equation}
The requirement of vanishing of $\gamma$ gives
$$\alpha_1^{(0)}=6d_2/d_1, \ \ \alpha_2^{(0)}= 3(d_3/d_1+12d_2^2/d_1^2+5d_0d_2/d_1).$$
So, one has
\begin{equation}\label{14}
y|_{\varepsilon=0}=g+6\frac{d_2}{d_1}g^2+3(\frac{d_3}{d_1}+
12\frac{d_2^2}{d_1^2}+5\frac{d_0d_2}{d_1})g^3+...
\end{equation}

If eq.(\ref{14}) is fulfilled then the anomalous dimension (and the beta function)
vanishes up to three loops and one has conformal invariance. Since we claim that
conformal invariance in this context is synonym to finiteness, let us check the
cancellation of UV divergences. As was explained above we will need eq.(\ref{13}) for
$\varepsilon\neq 0$
\begin{equation} \label{ar}
  y=g(1+\alpha_0^{(1)}\varepsilon+\alpha_0^{(2)}\varepsilon^2+...)
   +g^2(\alpha_1^{(0)}+\alpha_1^{(1)}\varepsilon+...)
  +g^3(\alpha_2^{(0)}+...).
\end{equation}
Notice that in the third order of PT the one should take into account all terms of the
double expansion with the total power of $g \cdot \varepsilon$ equal 3.

 Substituting
eq.(\ref{ar}) into (\ref{2}) one gets the remained coefficients
\begin{eqnarray}
\alpha_0^{(1)}&=& -2d_2/d_1^2,\ \ \alpha_0^{(2)}=\frac{2}{3d_1^2}(\frac{d_3}{d_1}+
6\frac{d_2^2}{d_1^2}-\frac{2a_2d_2}{3d_1^2}+15\frac{d_0d_2}{d_1}), \nonumber \\
\alpha_1^{(1)}&=&-\frac{2}{d_1}(\frac{d_3}{d_1}+
12\frac{d_2^2}{d_1^2}-\frac{2a_2d_2}{3d_1^2}+15\frac{d_0d_2}{d_1}). \label{coef}
\end{eqnarray}
With this choice of coefficients all the pole terms in $Z_2^{-1}$ cancel. Notice that if
$\alpha_1^{(0)}$ is responsible for the cancellation of the two-loop anomalous dimension,
both $\alpha_1^{(0)}$ and $\alpha_0^{(1)}$ are needed to cancel the $1/\varepsilon$ term
in two loops. They also cancel the $1/\varepsilon^2$ term in three loops. Indeed, taking
into account (\ref{ar}) it takes the form
$$\frac{1}{\varepsilon^2}: \ c_{32}|_{y=g}+y\frac{dc_{33}}{dy}|_{y=g}\alpha_0^{(1)}+
y\frac{dc_{22}}{dy}|_{y=g}\alpha_1^{(0)}=[-6d_2d_1+\frac
32d_1^3(-2)\frac{d_2}{d_1^2}+\frac 32d_1^26\frac{d_2}{d_1}]g^3=0 .$$

 Similarly,
$\alpha_2^{(0)}$ is needed to cancel the three loop anomalous dimension and all three
$\alpha_2^{(0)}$, $\alpha_1^{(1)}$ and $\alpha_0^{(2)}$ terms are used to cancel the
$1/\varepsilon$ term in three loops.

Consider now the chiral propagator (\ref{1}) and substitute our values of the
coefficients $d_{ij}$. One has for the singular part
\begin{eqnarray}\label{15}
  D_{B}(p^2,g_{B},h_{B})&=&1+\frac{d_1(g_B-y_B)}{\varepsilon}\frac{1}{(p^2)^\varepsilon}\\
  && \hspace*{-3cm}+
  \left(\frac{d_1^2(y_B-g_B)(4y_B-g_B)}{2\varepsilon^2}
  +\frac{d_2(g^2_B+g_By_B+y^2_B)}{\varepsilon}\right)
  \frac{1}{(p^2)^{2\varepsilon}} \nonumber \\ && \hspace*{-3cm}+
  \left(\frac{-d_1^3(y_B-g_B)(28y^2_B-20y_Bg_B+g^2_B)}{6\varepsilon^3}
  +\frac{-d_2d_1y_B(y^2_B+y_Bg_B+g^2_B)}{\varepsilon^2}\right. \nonumber \\
  &&\hspace*{-2.2cm} \left. \frac{-d_2d_1(y_B-g_B)(5y_B^2+3y_Bg_B+g_B^2)
  +1/3a_2d_1g^2_B(y_B-g_B)}{\varepsilon^2}\right. \nonumber \\
  &&\hspace*{-2.2cm} \left. \frac{-d_1^2d_0(y_B-g_B)(10y^2_B-2y_Bg_B-1/2g^2_B)}{\varepsilon^2}
   +\frac{d_3g^3_B}{\varepsilon}\right)\frac{1}{(p^2)^{3\varepsilon}} \nonumber.
\end{eqnarray}
To get the cancellation of divergences in each order of perturbation theory one again has
to choose the Yukawa coupling in a proper way in the form of $\varepsilon$ expansion
\begin{equation}\label{16}
  y_B=g_B\left(1+\alpha_{0}^{(1)} \varepsilon+\alpha_0^{(2)} \varepsilon^2+...\right).
\end{equation}
Substituting this expansion into (\ref{15}) and requiring the cancellation of
divergencies one gets for $\alpha_{0}^{(1)}$ and $\alpha_{0}^{(2)}$ the same values as
above (\ref{coef}). Contrary to the nullification of the anomalous dimension where the
cancellation takes place between the lower and higher orders of PT, here the cancellation
takes place within the same order between the higher and lower order pole terms. However,
these two procedures are related since the higher order poles are given via RG pole
equations by the lowest order expressions (see eq.(\ref{pole})). Notice that the
condition $y_B=g_B$ cancels the leading poles in all orders, the condition
$y_B=g_B(1+\alpha_0^{(1)}\varepsilon)$ cancels subleading poles in all orders, and the
condition $y_B=g_B(1+\alpha_0^{(1)}\varepsilon+\alpha_0^{(2)}\varepsilon^2)$ cancels the
subsubleading poles. In our case by the choice of $\alpha_0^{(1)}$ we cancel
$1/\varepsilon$ term in two loops and simultaneously $1/\varepsilon^2$ term in three
loops. The $\alpha_0^{(2)}$ term cancels the $1/\varepsilon$ term in three loops.  So,
one has
\begin{equation}\label{17}
y_B=g_B(1-2\frac{d_2}{d_1^2}\varepsilon+\frac{2}{3d_1^2}
(\frac{d_3}{d_1}+\frac{6d_2^2}{d_1^2}-\frac{2a_2d_2}{3d_1^2}
+\frac{15d_0d_2}{d_1})\varepsilon^2+...).
\end{equation}
If this conditions are satisfied then all divergences cancel and the theory is finite up
to three loops. Further loops require new terms in eq.(\ref{17}).

\section{Beta Deformed N=4 SYM Theory in 4 Loops}

Consider now the beta deformed ${\cal N}=4$ SYM theory. It is given by the
action~\cite{Z}
\begin{eqnarray}
  S&=& \int d^8z Tr\left(e^{-gV}\bar \Phi_i e^{gV} \Phi^i\right) +\frac{1}{2g^2}\int d^6z Tr
  (W^\alpha W_\alpha) \nonumber\\
  &&+ih\int d^6z Tr\left(q\Phi_1\Phi_2\Phi_3-\frac 1q \Phi_1\Phi_3\Phi_2\right)
  \nonumber\\
  &&+i\bar h\int d^6\bar z Tr\left(\frac{1}{\bar q} \bar\Phi_1 \bar \Phi_2\bar \Phi_3
  -\bar q \bar \Phi_1\bar \Phi_3\bar \Phi_2\right), \ \ q\equiv e^{i\pi\beta},
  \label{lag}
\end{eqnarray}
where the superfield strength tensor $W_\alpha= \bar D^2(e^{-gV}D_\alpha e^{gV})$ and
$\Phi_i$ with $i=1,2,3$ are the three chiral superfields of the original ${\cal N}=4$ SYM
theory in adjoint representation of the gauge group; $h$ and $\beta$ are complex numbers
and $g$ is the real gauge coupling constant. In the undeformed ${\cal N}=4$ SYM theory
one has $h=g$ and $q=1$.

In the present case it is useful to define the couplings
\begin{equation}\label{coup}
  h_1\equiv hq, \ \ \ h_2\equiv h/q, \ \ \ h_1^2\equiv h_1\bar h_1, \ \ \ h_2^2\equiv h_2\bar h_2.
\end{equation}

The goal is to study the conditions that in the planar limit (large $N$ of $SU(N)$)  the
couplings $h_1^2$ and $h_2^2$ have to satisfy in order to get conformal invariance of the
theory for complex values of $h$ and $\beta$. Explicit calculation gives the following
values for the coefficient functions of the renormalization constant $Z_2^{-1}$ in
notation of the previous section~\cite{Z} (For simplicity everywhere only the difference
between the beta deformed and undeformed ${\cal N}=4$ SYM theory is
considered~\cite{RSS})

\begin{equation}\label{cf}
  c_{nk} =F_{nk}(h_1^2,h_2^2,g^2)-(2g^{2})^{n}, \ \ \ \ n=1..3, \ k=1..3,
\end{equation}
where the functions $F_{nk}$ satisfy
$$F_{nk}(h_1^2+h_2^2=2g^{2})=(2g^{2})^{n}.$$
Eq.(\ref{cf}) can be also rewritten as
\begin{equation}\label{ci}
  c_{nk} =(h_1^2+h_2^2-2g^{2})P_{nk}(h_1^2,h_2^2,2g^{2}),\ \  n=1..3, k=1..3,
\end{equation}
where $P_{nk}$ is a homogeneous polynomial of the order $n-1$.

For $n=4$ one has
\begin{eqnarray}
c_{4i} &=& ( h_1^2+h_2^2-2g^{2})P_{4i}(h_1^2,h_2^2,2g^{2}),\ \ i\neq1 \label{c}\\
c_{41} &=&(h_1^2+h_2^2-2g^{2})P_{41}(h_1^2,h_2^2,2g^{2})+
G_{41}(h_{1}^{2},h_{2}^{2},2g^{2}).
\end{eqnarray}
where $G_{41}(h_{1}^{2},h_{2}^{2},2g^{2})$ is a homogeneous polynomial of the fourth
order that does not vanish at $h_1^2+h_2^2=2g^2$. The latter contribution comes from the
four loop chiral graph~\cite{Z} (see Fig.\ref{graph}). This graph has no divergent
subgraphs and, therefore, has only primitive divergence.
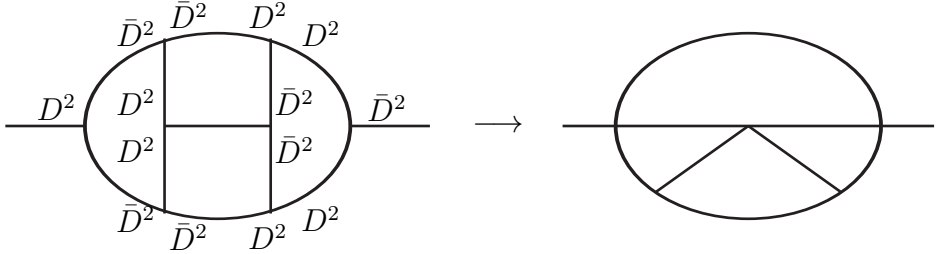
\begin{figure}[h]
  \centering
\begin{picture}(300,100)(200,-50)
\SetWidth{1} \Oval(255,0)(35,50)(360) \Line(275,33)(275,-33) \Line(235,33)(235,-33)
\Line(235,0)(275,0) \Line(175,0)(205,0) \Line(305,0)(335,0) \Text(360,0)[]{ $\mathbf
\longrightarrow$} \Oval(455,0)(35,50)(360)\Line(385,0)(525,0)\Line(455,0)(490,-25)
\Line(455,0)(420,-25)\Text(195,7)[]{$D^2$}\Text(225,9)[]{$D^2$}\Text(225,-9)[]{$D^2$}
\Text(275,42)[]{$D^2$}\Text(295,35)[]{$D^2$}\Text(275,-42)[]{$D^2$}\Text(295,-35)[]{$D^2$}
\Text(285,9)[]{$\bar D^2$}\Text(285,-9)[]{$\bar D^2$} \Text(245,42)[]{$\bar
D^2$}\Text(225,35)[]{$\bar D^2$}\Text(245,-42)[]{$\bar D^2$}\Text(225,-35)[]{$\bar D^2$}
\Text(320,7)[]{$\bar D^2$}
\end{picture}
\caption{The only relevant divergent planar supergraph and its scalar counterpart at four
loops}\label{graph}
\end{figure}

Explicit form of $c_{11}$ and $c_{41}$ is
\begin{eqnarray}
c_{11} &=& (-\frac{N}{(2\pi)^{2}})(h_1^2+h_2^2-2g^{2}) \doteq d_{1}(h_1+h_2^2-2g^{2})
\label{c2}\\ c_{41}
&=&\frac{5}{2}\zeta(5)\frac{N^{4}}{(2\pi)^{8}}[(h_1^2+h_2^2)^{4}-(2g^{2})^{4}
+(h_1^{2}-h_2^2)^4]\nonumber\\
& \doteq & d_{2}[(h_1^2+h_2^2)^{4}-(2g^{2})^{4}+(h_1^{2}-h_2^2)^4].
\end{eqnarray}
Hereafter the chiral-gauge $\bar{ \Phi } V \Phi$ contributions proportional to
$h_1^2+h_2^2-2g^{2}$ are omitted.

According to the recipe of the previous section one can now construct a conformal and
finite theory choosing the renormalized couplings in the form of a double series of the
fourth order
\begin{eqnarray}
h_{1}^{2}&=&g^{2}(a+\alpha_0^{(3)} \varepsilon^{3})+g^4\alpha_1^{(2)}\varepsilon^2
+g^6\alpha_2^{(1)}\varepsilon+g^{8}\alpha_3^{(0)} \ ,\nonumber\\
h_{2}^{2}&=&g^{2}(b+\beta_0^{(3)} \varepsilon^{3})+g^4\beta_1^{(2)}\varepsilon^2
+g^6\beta_2^{(1)}\varepsilon+g^{8}\beta_3^{(0)}. \label{expan}
\end{eqnarray}
Now, from the requirement of vanishing of anomalous dimension
$\gamma=c_{11}+2c_{21}+3c_{31}+4c_{41}=0$, one finds
\begin{eqnarray}
1\ loop: & &\ a+b=2, \label{van} \\
4\ loops: & & \ \alpha_3^{(0)}+\beta_3^{(0)}=\frac{-4\widehat G_{41}}{d_{1}g^8}=
\frac{-4(a-b)^{4}d_{2}}{d_{1}}, \nonumber
\end{eqnarray}
where hereafter $\widehat G_{41}$ means that one has to take $G_{41}$ at
$h_1^2+h_2^2=2g^2$.

To get $\alpha_0^{(3)}$ and $\beta_0^{(3)}$ one has to consider the bare propagator.
Since the only nontrivial graph giving contribution to $G_{41}$ has no divergent
subgraphs the essential singular part of the bare propagator is
$$D_{41}=-G_{41}.$$
Therefore, the condition for its cancellation is
$$\widehat P_{44}g^2(\alpha_0^{(3)}+\beta_0^{(3)})-\widehat G_{41}=0.$$
This gives
\begin{equation}\label{al}
\alpha_0^{(3)}+\beta_0^{(3)}=\frac{\widehat G_{41}}{\widehat P_{44}g^2}.
\end{equation}
The value of $\widehat P_{44}$ can be calculated from the pole equations: $\widehat
P_{44}=9d_1^4g^6$, so that
\begin{equation}\label{all}
\alpha_0^{(3)}+\beta_0^{(3)}=\frac{(a-b)^4d_2}{9d_1^4}.
\end{equation}

To reach total finiteness one can use the remaining parameters. From the requirement that
$ Z_{2}^{-1}=1$ in four loops one gets
\begin{equation}\label{first}
  \widehat G_{41}+d_1g^8(\alpha_3^{(0)}+\beta_3^{(0)})
  +\widehat P_{22}g^6(\alpha_2^{(1)}+\beta_2^{(1)})+
  \widehat P_{33}g^4(\alpha_1^{(2)}+\beta_1^{(2)})
  +\widehat P_{44}g^2(\alpha_0^{(3)}+\beta_0^{(3)})=0.
\end{equation}

This is one equation for two pairs of parameters. However, the same parameters are
responsible for the cancellation of the second order pole in five loops.  The fifth order
coefficients are
\begin{eqnarray}
c_{5i}& =& (h_1^2+h_2^2-2g^{2})P_{5i}(h_1^2,h_2^2,2g^{2}), \ \ \ \ \ \ \ \ \ i=3,4,5,
\label{five} \\
c_{5i}&=& (h_1^2+h_2^2-2g^{2})P_{5i}(h_1^2,h_2^2,2g^{2})+
G_{5i}(h_{1}^{2},h_{2}^{2},2g^{2}),\ i=1,2 \nonumber
\end{eqnarray}
Having in mind expansion (\ref{expan}) the second order pole takes the form
\begin{equation}\label{second}
\widehat G_{52}+\widehat P_{22}g^8(\alpha_3^{(0)}+\beta_3^{(0)}) + \widehat
P_{33}g^6(\alpha_2^{(1)}+\beta_2^{(1)})+ \widehat P_{44}g^4(\alpha_1^{(2)}+\beta_1^{(2)})
+ \widehat P_{55}g^2(\alpha_0^{(3)}+\beta_0^{(3)})=0.
\end{equation}
The coefficient functions $\widehat P_{22},\widehat P_{33}, \widehat P_{44}, \widehat
P_{55}$ and $\widehat G_{52}$ can be found from the pole equations that gives
$$
\widehat P_{22}=3d_{1}^2g^2, \ \widehat P_{33}=6d_{1}^3g^4,\ \widehat
P_{44}=9d_{1}^4g^6,\ \widehat P_{55}=\frac{54}{5}d_{1}^5g^8, \ \ \ \widehat
G_{52}=\frac{24}{5}d_1\widehat G_{41}g^2.
$$
Substituting these values  into (\ref{first},\ref{second}) and taking into account
eqs.(\ref{van},\ref{all}) one gets
\begin{eqnarray}
\alpha_1^{(2)}+\beta_1^{(2)}&=&-\frac 23\frac{(a-b)^4d_2}{d_1^3}, \nonumber \\
\alpha_2^{(1)}+\beta_2^{(1)}&=&2\frac{(a-b)^4d_2}{d_1^2}. \label{rem}
\end{eqnarray}

Provided eqs.(\ref{van},\ref{all},\ref{rem}) and (\ref{expan}) are satisfied one has
totally consistent finite and conformally invariant theory (up to four loops)
parameterized by two parameters $a$ and $b$ related by one condition $a+b=2$. Apparently
the mechanism will work in any loop order irrespectively of the explicit form of
divergent terms. Looking back to the analysis of divergent structures in Ref.\cite{Z} one
finds that new chiral graphs always give contribution  proportional to $(h_1^2-h_2^2)^4$,
so that the compensating terms of expansion will be always proportional to $(a-b)^4$ as
above.

In Ref.\cite{Z,Z1}  it was claimed that the only reliable solution is $a=b=1$. Otherwise
one can not reach both the finiteness and conformal invariance simultaneously. We see
that this statement is a result of mistreatment of dimensional regularization (reduction)
in the process of cancellation of divergences: the authors of \cite{Z,Z1} considered only
the one fold expansion instead of two fold one (\ref{expan}). For the correct
implementation of the procedure $a$ is arbitrary and $b=2-a$. In fact, as one can see
above, the requirement of cancellation of divergences always defines only the sum of
$\alpha$'s and $\beta$'s, thus allowing the whole family of solutions
\begin{eqnarray}
&&h_{1}^{2}+h_2^2=\bar h h (\bar q q+ 1/\bar q q) \label{res} \\
&& =g^{2}\left\{2+\frac{5}{18}\zeta_5\delta^4\varepsilon^{3}
+\frac{5}{3}\zeta_5\delta^4(\frac{g^2N}{4\pi^2})\varepsilon^2
+5\zeta_5\delta^4(\frac{g^2N}{4\pi^2})^2\varepsilon+10\zeta_5\delta^4
(\frac{g^2N}{4\pi^2})^3+...\right\}, \nonumber
\end{eqnarray}
where we denoted $a-b\equiv\delta$.  For the bare couplings one has
\begin{equation}\label{res2}
h_{1}^{2}|_B+h_2^2|_B=g^{2}_B\left\{2+\frac{5}{18}\zeta_5\delta^4\varepsilon^{3}+...\right\}.
\end{equation}
This permits, in particular, the value of $|q|\neq 1$, thus allowing one to obtain a
complex deformation of the ${\cal N}=4$ SYM theory with arbitrary complex $\beta$.

\section{Conclusion}

We conclude that properly treated $\beta$ deformed ${\cal N}=4$ SYM theory can be made
{\it simultaneously} conformal invariant and finite since these two requirements are {\it
identical}. This can be achieved by adjusting the Yukawa couplings order by order in PT.
In the framework of dimensional regularization (reduction) this requires the double
series over the gauge coupling $g$ and the parameter of dimensional regularization
$\varepsilon$. For the bare coupling, on the contrary, only the one fold series over
$\varepsilon$ is enough. The whole procedure depends on regularization (for bare
quantities) and renormalization scheme (for the renormalized ones). In the other
regularization techniques it looks differently but the main conclusion remains the same.

The analysed $\beta$ deformed SYM theory represents the whole class of conformal ${\cal
N}=1$ SYM theories in four dimensions. They can be constructed by the same mechanism of
adjustment of the corresponding Yukawa couplings. This adjustment has to be done order by
order in PT.  At the moment there is no any theory (except for ${\cal N}=4$ and ${\cal
N}=2$ SYM ones) for which the all loop solution is known. These theories may as well have
a dual description in the framework of supergravities within the AdS/CFT correspondence,
though the proper backgrounds are not found but few steps in this direction have been
made (see for example \cite{betadual1},~\cite{betadual2}).

\section*{Acknowledgements}

Financial support from RFBR grant \# 05-02-17603 and grant of the Ministry of Education
and Science of the Russian Federation \# 5362.2006.2 is kindly acknowledged. We are
grateful to E. Sokatchev for attracting our attention to the subject and valuable
discussions.

\end{document}